\documentstyle[epsf,psfig,aps]{revtex}

\newcommand{\bfm}[1]{\mbox{\boldmath $#1$}}

\newcommand{\la}{\langle}

\newcommand{\ra}{\rangle}

\newcommand{\rta}{\rightarrow}

\newcommand{\nIs}{n_{1\s}}

\newcommand{\nJs}{n_{2\s}}

\newcommand{\R}{\Rv}

\newcommand{\Rv}{\bfm{R}}

\newcommand{\rv}{\bfm{r}}

\newcommand{\s}{\sigma}

\newcommand{\beq}{\begin{equation}}

\newcommand{\eeq}{\end{equation}}

\begin{document}

\centerline{\bf DISTANCE-DEPENDING ELECTRON-PHONON INTERACTIONS}
\vspace*{0.035truein}
\centerline{\bf  FROM ONE- AND TWO-BODY ELECTRONIC TERMS IN A DIMER.}
\vspace*{0.37truein} \centerline{M. ACQUARONE}
\vspace*{0.015truein} \centerline{\footnotesize\it
C.N.R.-G.N.S.M., Unita' I.N.F.M., Dipartimento di
Fisica,Universita' di Parma} \baselineskip=10pt
\centerline{\footnotesize\it Parma, I-43100, Italy} \vspace*{10pt}
\centerline{\normalsize and} \vspace*{10pt} \centerline{C. NOCE}
\vspace*{0.015truein} \centerline{\footnotesize\it Unit\`a
I.N.F.M., Dipartimento di Scienze Fisiche ''E.R. Caianiello'',
Universit\`a di Salerno,} \baselineskip=10pt \centerline{\it
Baronissi, I-84081, Italy} \vspace*{0.225truein}

\vspace*{0.21truein}

{For a dimer with a non-degenerate orbital built from atomic wave
functions of Gaussian shape we evaluate all the  electron-phonon
couplings derived from the one-body and two-body electronic
interactions, considering
 both the adiabatic and extreme non-adiabatic
limit. Not only the  values
 of the coupling parameters in the two limits, but also the
 expressions of the corresponding terms in the
Hamiltonian differ.
 Depending on the distance between the dimer ions,
some of the two-body couplings are comparable, or even larger than
the one-body ones.}{}{}



\section{The Model.}    

\vspace*{-0.5pt}

\noindent

 In a general two-site electron-phonon Hamiltonian
$H=H_{el}+H_{ph}+H_{el-ph}  \label{ham1}$ the interacting term
$H_{el-ph}$ originates from developing $H_{el}$ to first order in
the ion displacements $u_i\quad (i=1,2)$, where, in standard
notation for a non-degenerate orbital:
\[
H_{el}=\epsilon \sum_{\sigma }(n_{1\sigma
}+n_{2\sigma})+\sum_{\sigma} [t+X(n_{1-\sigma
}+n_{2-\sigma})](c_{1\sigma }^{\dagger }c_{2\sigma }+H.c.)
+U(n_{1\uparrow}n_{1\downarrow}+n_{2\uparrow}n_{2\downarrow})
\]
\begin{equation}
+(V-J/2)n_1n_2-2J\left[S^z_1S^z_2+{{1}\over{2}}(S^+_1S^-_2+H.c.)\right]
+P(c^{\dagger}_{1\uparrow}c^{\dagger}_{1\downarrow}
c^{}_{2\downarrow}c^{}_{2\uparrow}+H.c.).
 \label{helbare}
\end{equation}
 We shall develop both the one-($\epsilon,t$) and two-body($X,U,V,J_z=J_{xy}=P$)
 electron interaction parameters, evaluated as in Ref.1, by assuming
 a non-degenerate orbital described
by Wannier functions built from atomic orbitals of Gaussian shape.
We associate to each site a Gaussian atomic-like orbital
$\phi_i(\rv-\R_i)$, with the ions centered at  the positions
$\R_i\equiv(\pm a/2+u_i,0,0)\quad (i=1,2)$ . By defining
$N\equiv\left (2/ \pi\right)^{3/4}\Gamma^{3/2}$, they read:
$\phi_i(\rv)= N\exp\left\{-\Gamma^2\left[\left(x\pm
a/2-u_i\right)^2+y^2+z^2\right]\right\}$.
 Then the Wannier functions
$\Psi_1, \Psi_2$ can be written as:
\[
\Psi_1(\rv)=A(S) \phi_1(\rv)+ B(S) \phi_2(\rv) \qquad
\Psi_2(\rv)=B(S) \phi_1(\rv) +A(S)\phi_2(\rv)
\]
\beq
A(S)=[(1+S)^{-1/2}+(1-S)^{-1/2}]/2 \qquad B(S)=[(1+S)^{-1/2}-(1-S)^{-1/2}]/2
\eeq
 with $S(u) \equiv \la \phi_1|\phi_2\ra =\exp[-\Gamma^2(a+u)^2/2]$, and $u=u_2-u_1$.
 To make clear our method of calculation, it is convenient to explicitate, as an example,
 the one-body local electronic energy:
\beq \epsilon^{(i)}=\int\Psi_i^{\ast}(\rv,\R_1,\R_2)
\left[-{{\hbar^2}\over{2m}}\nabla^2+V_1(\rv-\R_1)+V_2(\rv-\R_2)\right]
\Psi_i^{~}(\rv,\R_1,\R_2) d^3\rv \eeq
 where the potentials originating from the ion cores at
the displaced positions $\R_1$ and $\R_2$ are: \beq V_i\equiv
V(\rv-\R_i)=-e^2Z\left[\left(x\pm a/2- u_i
\right)^2+y^2+z^2\right]^{-1/2}\qquad (i=1,2) \eeq with $-e$ the
electron charge, and $+Ze$ the charge of the ion core. The local
energy  $\epsilon$ (actually site-independent) can be decomposed
into three terms, corresponding to the contributions from the
kinetic operator ($\epsilon_{\nabla}^{(i)}$) and from each one of
the  ionic potentials
($\epsilon_{V_1}^{(i)},\epsilon_{V_2}^{(i)}$, respectively).
 A similar decomposition holds for the hopping amplitude $t$. \par
We need to  distinguish between the adiabatic
 and the non-adiabatic
limit in evaluating the electron-phonon interactions, because the
integrals have different kernels in the two cases. Indeed, in the
adiabatic limit the  displacements affect both the potentials $
V(\rv-\R_i)$ and the  Wannier functions
$\Psi_i^{~}(\rv,\R_1,\R_2)$ , expressing the requisite that the
electronic charge distribution adjusts itself instantaneously at
the position of the ions. We shall schematize the opposite
situation, where the electrons are slower than the ions, as
realized by  the electronic charge distribution staying centred
around the undisplaced ion position, while the potentials are
centred on the displaced ions.
 We shall call this the extreme anti-adiabatic limit. In the literature\cite{ashkenazi}
 the two limits are also named from, respectively, Fr\"{o}lich and Bloch.\par





\section{Coupling terms in the anti-adiabatic limit.}
\noindent In the anti-adiabatic limit  $\epsilon^{~}_{\nabla}$
does not change, therefore no electron-phonon coupling originates
from it. The coupling terms derived from the two-body interactions
are also identically vanishing in this limit, because they involve
the Wannier functions and the inter-electronic Coulomb potential
 which are both insensitive to the displacements of the ions.
The only non-vanishing electron-phonon non-adiabatic interactions
arise from the variation of  the potential contributions to
$\epsilon$ and $t$. Let us now succinctly describe their
evaluation. Full details are given in Ref.3. The perturbation of
$\epsilon^{(i)}_{V_i}$ originates a term in the Hamiltonian
connecting the local charge with the local deformation: $\sum_\s
(g_0^{(1)}\nIs u_1+g_0^{(2)}\nJs u_2)$ . It has  a formal
similarity with the Holstein coupling term\cite{holstein}, but it
has a physically different origin, as Holstein\cite{holstein}
 considered electrons moving along  a chain of fixed spacing
 with vibrating diatomic molecules at its nodes.
On site 1, its explicit evaluation\cite{macanio} yields: \beq
g_0^{(1)}={{2\Gamma\sqrt{2/\pi} } \over {a}} \left\{B^2
\left[F_0(2a^2\Gamma^2)  -S^4\right] +4ABS \left[
F_0\left({{a^2\Gamma^2}\over{2}}\right)-S \right] \right\}.
\label{hol27c} \eeq \noindent where $F_0(x)=x^{-1/2}Erf(x^{1/2})$.
As, under site permutation, $a\rightarrow -a$,
$g_0^{(1)}=-g_0^{(2)}$, as might have been anticipated  by
considering that, for equal charges $n_1=n_2$ and displacement
amplitudes, with ${\bf e}_{12}\equiv (\Rv_2-\Rv_1)/a$, the
energies $g_0^{(1)}n_1{\bf u}_1\bullet{\bf e}_{12}$ and
$g_0^{(2)}n_2{\bf u}_2\bullet{\bf e}_{12} $ on both sites
coincide. Now symmetry requires $u_1=-u_2$ (a constraint which
does not hold for the original Holstein model\cite{holstein}) from
which $g_0^{(2)}=-g_0^{(1)}$ follows. The contribution to the
Hamiltonian then reads:
$g_0^{(1)}\sum_{\sigma}(n_{1\sigma}^{~}u_1^{~}-n_{2\sigma}^{~}u_2^{~})$
The ``crystal-field'' coupling term $g_{cf}^{(ij)}$ expresses the
change in the energy $\epsilon_{V_j}^{(i)}$ . To establish the
form of this term, let us consider site $1$,
 with charge $n_{1}$.
Its energy, after a displacement ${\bf u}_{2}$ of the ion on site
$2$, changes by an amount \nobreak{$E^{(1)}=g_{cf}^{(12)}n_{1}$
${\bf u}_{2}\bullet {\bf e}_{12}$}. \ This can be considered as
the quantity measured by an observer sitting on ion $1$ and
watching the ion $2$ \ moved by ${\bf u}_{2}$ .\ The equivalent
measurement
done by  an observer \ on ion $2$ watching the ion $1$ displaced by ${\bf u}%
_{1}$,\ yields $E^{(2)}=g_{cf}^{(21)}n_{2}{\bf u}_{1}\bullet {\bf
e}_{12}$.  Assuming $n_1=n_2,and u_1=-u_2$ one must have
$E^{(1)}=E^{(2)}$, implying $g_{cf}^{(12)}=-g_{cf}^{(21)}$.
 Therefore for the dimer as a whole one writes this term as
$g_{cf}^{{12}}\sum_{\sigma}(n_{1\sigma}u_{2}-n_{2\sigma}u_{1})$.
The explicit evaluation\cite{macanio} for site $1$ yields:

 \beq
g^{(12)}_{cf}=-2A\left({{\Gamma}\over{a}}\right)\sqrt{{{2}\over{\pi}}}
\left[AF_0(2a^2\Gamma^2)+4BSF_0(a^2\Gamma^2/2)-4BS^2-AS^4\right].
\label{g1cfnadexpl} \eeq

From Eq.\ref{g1cfnadexpl} and the change of sign of $a$ under site
permutation,
 $g^{(12)}=-g^{(21)}$ follows.\par
The Su-Schrieffer-Heeger\cite{SSH} ($SSH$) interaction
$\gamma^{(12)}$, characterizing
 the $SSH$ Hamiltonian $ H_{SSH}\equiv \gamma^{(12)}\sum_\sigma (c_{1\sigma}^\dagger
c^{~}_{2\sigma}+c_{2\sigma}^\dagger c^{~}_{1\sigma})(u_2-u_1)$,
 is due to the modulation of the hopping amplitude $t$.
To preserve the invariance of $H_{SSH}$ under site permutation,
$\gamma^{(12)}=-\gamma^{(21)}$ (see e.g refs.6 and 7). Indeed, its
explicit expression\cite{macanio} is:

\[
\gamma^{(12)}=4\sqrt{{{2}\over{\pi}}}\left({{\Gamma}\over{a}}\right)
\left\{{{AB}\over{2}}\left[S_0^4-(1-4a^2\Gamma^2)F_0(2a^2\Gamma^2)\right]
\right\}
\]
\beq +4\sqrt{{{2}\over{\pi}}}\left({{\Gamma}\over{a}}\right)
\left\{(A^2+B^2)S_0\left[S_0-F_0(a^2\Gamma^2/2)\right]\right\}.
\label{ga12nadexpl} \eeq

Under site permutation $a\rta -a$ and $A\rta B$ so that
$\gamma^{(12)}=-\gamma^{(21)}$ as expected.

In conclusion, in the non-adiabatic limit the complete
electron-phonon Hamiltonian is given by:
\[
H^{na}_{ep}= g_0^{(1)}\sum_\s (\nIs u_1-\nJs u_2)+
g_{cf}^{(12)}\sum_{\sigma}(n_{1\sigma}u_{2}-n_{2\sigma}u_{1})
\]
\beq +\gamma^{(12)}\sum_\sigma (c_{1\sigma}^\dagger
c^{~}_{2\sigma}+c_{2\sigma}^\dagger c^{~}_{1\sigma})(u_2-u_1)
\label{Hnonad} \eeq

\section{Coupling terms in the adiabatic limit.}

\noindent
 In the explicit expression of the different electronic
 interactions in the adiabatically displaced state, $u$
invariably enters in the combination $a+u$. So, to obtain the
corresponding couplings,
 one can simply take the derivative with respect to $a$ of the
parameters in Eq.\ref{helbare} as evaluated in Ref.1.

There is some confusion in the literature about the correct form
of the electron-phonon Hamiltonian obtained, in the adiabatic
limit, from the variation of the local energy $\epsilon$,
therefore we shall devote some space to clarify this point. In
this limit, $u_i\ne 0$ enters  both the charge distributions and
the potentials.
 The `` crystal field'' interaction couples the
charge on site $i$ to the {\it relative} position of site $j$
through the modification of both the kinetic and the potential
contributions to $\epsilon^{(i)}$.

Let's place the origin of the $x$-coordinate onto one of the
displaced ions, at $\Rv_i$, say. One has then to take into account
the {\it relative} displacement of the ions. In the adiabatic
limit, therefore, the overall $\epsilon$-derived electron-phonon
coupling term in the Hamiltonian is: $g_{\epsilon}^{(12)}\sum_\s
n_{1\s}(u_2-u_1) +g_{\epsilon}^{(21)}\sum_\s n_{2\s}(u_1-u_2)$.
 As $g_{\epsilon}^{(12)}=-g_{\epsilon}^{(21)}$
 we can  write the total adiabatic
contribution from local energy terms to the electron-phonon
Hamiltonian as $H^{\epsilon}_{ep}=
g_{\epsilon}^{(12)}\sum_\s(\nJs+\nIs)(u_2-u_1)$. A Hamiltonian of
this form was used in the papers of  Ref.8, while those of Ref.9
proposed hamiltonianns incompatible with our results. After
including the $SSH$ term, the complete one-body electron-phonon
Hamiltonian in the adiabatic limit has therefore the
form\cite{noi2}:

\beq H^{ad}_{ep}=
g_{\epsilon}^{(12)}\sum_{\sigma}(n_{1\sigma}+n_{2\sigma})(u_{2}-u_{1})
 +\gamma^{(12)}\sum_\sigma (c_{1\sigma}^\dagger
c^{~}_{2\sigma}+c_{2\sigma}^\dagger c^{~}_{1\sigma})(u_2-u_1).
\label{Hadiab} \eeq

Coming now to the explicit expressions of the one-body coupling
parameters, let us write for conveniency
 $g^{(12)}_{\epsilon}\equiv g_{\nabla}^{(12)}+ g_{V}^{(12)}$
and $\gamma^{(12)}\equiv
\gamma^{(12)}_{\nabla}+\gamma^{(12)}_{V}$. We obtain:
\[
 g^{(12)}_{\nabla}=
-{{\hbar^2}\over{2m}}\left[{{a\Gamma^4S^2}\over{(1-S^2)^2}}\right]
\left[2\left(1-a^2\Gamma^2-S^2\right)\right]
\]
\[
 g^{(12)}_{V}=-Ze^2\left(2\Gamma\sqrt{{{2}\over{\pi}}}\right)
\left[{{\partial (A^2+B^2)}\over{\partial
u}}+{{4ABS^2+(A^2+B^2)S^4}\over{a}}\right]
\]
\beq -Ze^2\left(2\Gamma\sqrt{{{2}\over{\pi}}}\right)
\left\{F_0(2a^2\Gamma^2)\left[{{\partial (A^2+B^2)}\over{\partial
u}}- {{(A^2+B^2)}\over{a}}\right]+ (4S)\left[{{\partial
AB}\over{\partial u}}
-{{AB}\over{a}}\left(1+a^2\Gamma^2)\right)\right]\right\}.
 \label{geps}
\eeq

\[
\gamma^{(12)}_{\nabla}=
{{\hbar^2}\over{2m}}\left[{{aS\Gamma^4}\over{(1-S^2)^2}}\right]
\left[2(1-S^2)-a^2\Gamma^2(1+S^2)\right]
\]
\[
\gamma^{(12)}_{V}=
-Ze^2\left(4\Gamma\sqrt{{{2}\over{\pi}}}\right)\Bigg\{
\left[{{\partial AB}\over{\partial u}}
+{{A^2+B^2}\over{a}}S^2+{{AB}\over{a}}S^4\right] +
\left[{{\partial AB}\over{\partial u}}
-{{AB}\over{a}}\right]F_0(2a^2\Gamma^2)
\]

\begin{equation}
+S\left[{{\partial (A^2+B^2)}\over{\partial u}}
-{{A^2+B^2}\over{a}}(1+a^2\Gamma^2)\right]F_0(a^2\Gamma^2/2)\Bigg\}.
\label{SSHV}
\end{equation}
Notice that, as the partial derivatives are linear in $a$, they
 change sign under site permutation.
\par
Also the two-body electronic interactions $U,V,J(=P),X$ of Eq. (1)
 give rise to electron-phonon couplings, all of the form:
\beq H_{Y}^{ep}= \left({{dY}\over{da}}\right)
F(c^{\dagger}_{i\sigma},c^{~}_{j\sigma}) (u_j-u_i)\quad (i,j=1,2)
\eeq
 where $Y=U,V,X,J$ and
$F(c^{\dagger}_{i\sigma},c^{~}_{j\sigma})$ is the function of
Fermi operators representing the two-body interaction whose
amplitude is $Y$. From the results of Ref.1, their evaluation is
trivial. Notice that, as $U(a)=J(a)+e^2\Gamma/\sqrt{\pi}$, then
$dU/da=dJ/da=dP/da$. We list below their explicit expressions:
\[
{{dX}\over{da}}=-e^2{{\Gamma}\over{\sqrt{\pi}}}
\left[\left(-a\Gamma S\right)
 \frac{\left( 1+3S^{2}\right) }{(1-S^{2})^{3}}\right]
\left[1+2S^{2}+F_{0}\left( a^{2}\Gamma ^{2}\right)
 -2(1+S^{2})F_{0}\left( \frac{a^{2}\Gamma ^{2}}{4}\right) \right]
\]
\[
-e^{2}\frac{\Gamma }{\sqrt{\pi }}\left[ \frac{S/a}{(1-S^{2})^{2}}\right]
\left\{ 4a^{2}\Gamma ^{2}S^{2}\left[ F_{0}\left( \frac{a^{2}\Gamma ^{2}}{4}%
\right) -1\right] +S^{2}-F_{0}\left( a^{2}\Gamma ^{2}\right) \right\}
\]
\begin{equation}
-e^{2}\frac{\Gamma }{\sqrt{\pi }}\left[
\frac{S/a}{(1-S^{2})^{2}}\right] \left\{ 2(1+S^{2}) \left[
F_{0}\left( \frac{a^{2}\Gamma ^{2}}{4}\right) -\sqrt{S}\right]
\right\},   \label{dXda}
\end{equation}
\[
\frac{dU}{da}=
e^{2}\frac{\Gamma }{\sqrt{\pi }}\left[ \frac{-4a\Gamma
^{2}S^{2}}{(1-S^{2})^{3}}\right] \left[ 2-S^{2}+2S^{4}+S^{2}F_{0}\left(
a^{2}\Gamma ^{2}\right) -4S^{2}F_{0}\left( \frac{a^{2}\Gamma ^{2}}{4}\right)
\right]
\]
\[
+e^{2}\frac{\Gamma }{\sqrt{\pi }}\left[ \frac{S^{2}/a}{(1-S^{2})^{2}}\right] %
\Bigg\{2a^{2}\Gamma ^{2}\left[ 1-4S^{2}-F_{0}\left( a^{2}\Gamma ^{2}\right)
+4F_{0}\left( \frac{a^{2}\Gamma ^{2}}{4}\right) \right]
\]
\begin{equation}
+S^{2}-F_{0}\left( a^{2}\Gamma ^{2}\right) +4\left[ F_{0}\left( \frac{%
a^{2}\Gamma ^{2}}{4}\right) -\sqrt{S}\right] \Bigg\},
\label{dUda}
\end{equation}

\[
\frac{d(V-J_z/2)}{da}=
\]
\[
e^{2}\frac{\Gamma }{\sqrt{\pi }}\left[- \frac{4a\Gamma
^{2}S^{2}}{(1-S^{2})^{3}}\right] \left[
3-S^2-8S^4-(7-5S^2)F_0(a^2\Gamma^2)
-4(1-3S^2)F_0\left({{a^2\Gamma^2}\over{4}}\right)\right]
\]
\[
+e^{2}\frac{\Gamma }{\sqrt{\pi }}\left[ \frac{1/a}{(1-S^{2})^{2}}\right] %
\Bigg\{2a^{2}\Gamma ^{2}S^{2}\left[ -1-4S^{2}+F_{0}\left(
a^{2}\Gamma ^{2}\right) +4F_{0}\left( \frac{a^{2}\Gamma
^{2}}{4}\right) \right]
\]

\begin{equation}
+2S^2+{{3}\over{2}}S^4-\left(2-{{3}\over{2}}S^2\right)F_0(a^2\Gamma^2)
+2S^{2} \left[ F_{0}\left( \frac{a^{2}\Gamma
^{2}}{4}\right)-\sqrt{S} \right] \Bigg\}, \label{dVda}
\end{equation}


Fig.1 and 2 present the values of the one-body coupling constant
for, respectively, the non- adiabatic, and the adiabatic limit,
evaluated by assuming for the shape-controlling parameter  of the
Wannier functions $\Gamma$ the typical\cite{acquarone} value
$1.$\AA$^{-1}$.

\begin{figure}
\centerline{\psfig{figure=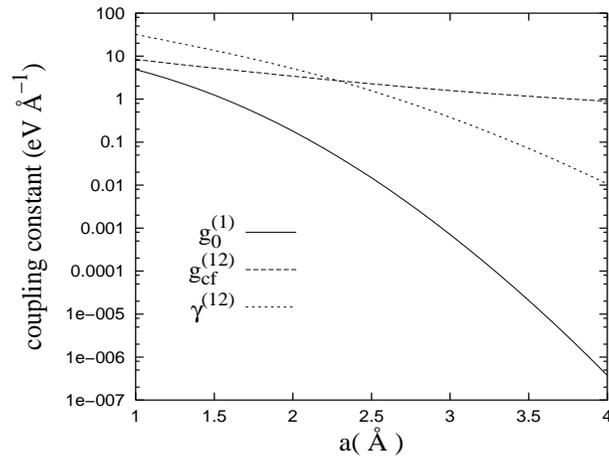,height=6cm,width=8cm}}


\caption {Non-adiabatic coupling constants
$g_0^{(1)},g_{cf}^{(12)},\gamma^{(12)}$ (in eV\AA$^{-1}$) versus
the dimer length (in \AA), evaluated assuming $\Gamma=1.0
\AA^{-1}$.} \label{fig1}
\end{figure}

 The  most unexpected result concerns $g_{cf}$. While usually neglected in
the literature\cite{barisic} on metallic systems, this coupling
has been recognized as relevant to polar materials\cite{barisic2}.
We find indeed that, when $\Gamma=1.0$\AA$^{-1}$, $g_{cf}$ is
larger than $g_0$ for any $a$, and it becomes the largest
parameter for $a>2.2$\AA. For small $a$, the $SSH$ coupling is the
largest.

\begin{figure}
\centerline{\psfig{figure=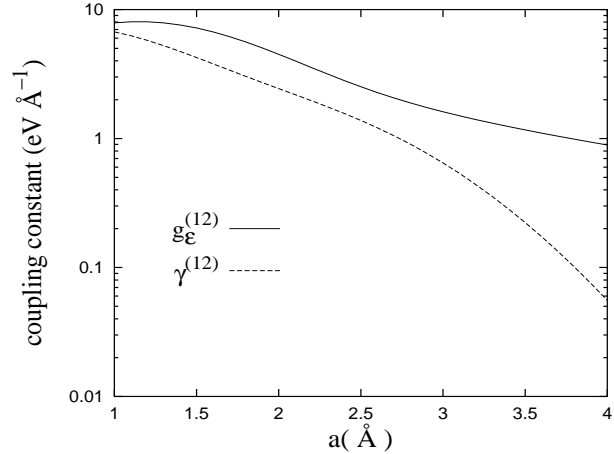,height=6cm,width=8cm}}
\caption {Adiabatic coupling constants $g_{\epsilon}^{(12)},
\gamma^{(12)}$ (in eV$\AA^{-1}$) versus the dimer length (in \AA),
for $\Gamma=1.0$ \AA$^{-1}$.} \label{fig2}
\end{figure}

Fig.2 for the adiabatic case shows that $g_{\epsilon}^{(12)}$ is
always larger than the $SSH$ interaction $\gamma^{(12)}$, and
particularly for large $a$ there is an order of magnitude
difference between them.

\begin{figure}
\centerline{\psfig{figure=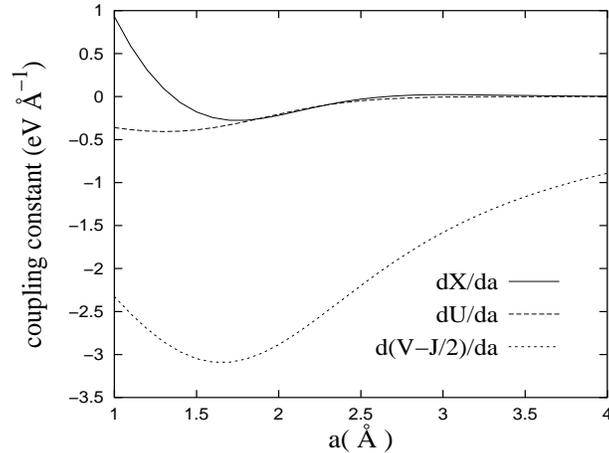,height=6cm,width=8cm}}
\caption {Adiabatic coupling constants from two-body interactions
 $dX/da$, $dU/da$, $d(V-J/2)/da$ (in eV\AA$^{-1}$) versus
the dimer length (in \AA), for $\Gamma=1.0$ \AA$^{-1}$.}
\label{fig3}
\end{figure}

The case of $\Gamma=2.\AA^{-1}$ (discussed in ref.3) shows that,
the more localized are the orbitals, the more relevant is the role
of $g_{cf}$ in relation to the other admissible couplings.

Fig.3 shows the couplings derived from the two-body electronic
interactions for the same parameters as Fig.2. In general, their
values are much smaller than those of $g_{\epsilon}$ and
$\gamma^{(12)}$, with the possible exception of $d(V-J/2)/da$.
Indeed, that coupling arises from a physical mechanism not very
different from the one originating $g^{(i)}_{V_j}$, i.e. the
vibration of the charge on site $j$ as felt by site $i$. Similarly
to $g_{\epsilon}$ also $d(V-J_z/2)/da$ decreases slowly with $a$,
so that for large $a$ those two are the only relevant couplings.
Such interactions in the lattice have been recently discussed in
ref.12, while their effects in the optical spectra have been
treated in ref.13.

\section{Conclusions}

\noindent We have presented the analytical evaluation of the electron-phonon
coupling parameters derived from both one- and two-body electronic interactions
 in a model of a dimer with a non-degenerate orbital
built from atomic orbital of Gaussian shape. We have shown that
the coupling terms in the adiabatic and the anti-adiabatic limits differ qualitatively.\par
The evaluation of the coupling terms originating from the two-body
electronic interactions shows that at least the one
generated by the Coulomb repulsion between the charges on
different sites, is comparable to, or even larger than, the couplings
derived from the one-body interactions. The quantitative
results for  the coupling parameters, even if agreeing in order of
magnitude with some estimates from experimental data\cite{hlubina}
 are obviously model-depending. However, their ratios
should be more close to the reality. In particular, the obtained
values of the coupling terms, when compared to the values of
the electronic interactions resulting from the same Wannier
functions\cite{acquarone}
 suggests that, for dimer lengths comparable to the lattice parameters
in high temperature superconductors and colossal magnetoresistance
materials, at most $dU/da$ and $dX/da$ can be safely dropped,
while neglecting {\it any} of the other electron-phonon
interactions is a questionable approximation.

 {Acknowledgements}
\noindent It is a pleasure to thank J.R. Iglesias, M. A.
Gusm\~{a}o, M. Cococcioni, A. Alexandrov, and particularly A.A.
Aligia and A. Painelli, for critical discussions and comments.
This work was supported by

I.N.F.M. and by MURST 1997 co-funded project "Magnetic Polarons in
Manganites".\par


\begin{thebibliography}{99}
\bibitem{noi2} The Hamiltonians of Eqs. \ref{Hnonad} and \ref{Hadiab} differ
 from the
corresponding results in Ref.3, because here we correct some
 mistakes which we found in that paper after it was published.
\end{thebibliography}
\end{document}